\documentclass{PoS}

\usepackage{amsmath}
\usepackage{multicol}
\usepackage{mathtools}
\usepackage{textpos}

\title{Charmonium Potentials at Non-Zero Temperature}

\ShortTitle{Charmonium Potentials at Non-Zero Temperature}

\author{\speaker{P. Wynne M. Evans}\\
        Department of Physics, College of Science, Swansea University, Swansea, United Kingdom\\
        E-mail: \email{pyevans@swan.ac.uk}}
\author{Chris Allton\\
        Department of Physics, College of Science, Swansea University, Swansea, United Kingdom\\
        E-mail: \email{c.allton@swan.ac.uk}}
\author{Pietro Giudice\\
        Institut f\"ur Theoretische Physik, Universit\"at M\"unster, M\"unster, Germany\\
        E-mail: \email{p.giudice@uni-muenster.de}}
\author{Jon-Ivar Skullerud\\ 
        Department of Mathematical Physics, National University of Ireland Maynooth, Maynooth, County Kildare, Ireland\\
        E-mail: \email{jonivar@thphys.nuim.ie}}

\abstract{The charmonium potential at non-zero temperature has been studied using gauge configurations with anisotropic lattices and 2+1 dynamical flavors of light sea quarks. We use the HAL QCD time-dependent method developed for the study of nucleon-nucleon potentials. To serve as input, local-extended charmonium correlators were calculated. The results are consistent with the expectation that the potential between heavy quarks should become deconfining at high temperatures.}

\FullConference{31st International Symposium on Lattice Field Theory LATTICE 2013\\
                 July 29 - August 3, 2013\\
                 Mainz, Germany}

\begin{document}
\section{Introduction}
\label{int}
In 1986 Matsui and Satz detailed in their seminal paper how $J/\psi$ suppression could act as a signal for quark-gluon plasma (QGP) formation in heavy ion collisions \cite{Matsui:1986dk}. There is now a significant body of experimental evidence for $J/\psi$ suppression and the potential picture provides an explanatory mechanism for this observation  \cite{Abreu:2000ni}. Quarkonium suppression has also been observed in bottomonium states \cite{Dahms:2013eaa}. Sequential suppression of heavy quarkonium states has been suggested as a means to gauge the temperature produced in heavy ion collisions. The motivation for calculating the charmonium potential, especially from first principles, therefore lies in the capacity for accurate heavy quarkonium potentials to become valuable QGP diagnostic aids.

In this work the HAL QCD \textit{time-dependent} method, described in Section \ref{app}, was used to calculate the charmonium potential \cite{Aoki:2012tk}. This method was used in \cite{Evans:2013yva} but is distinct from the HAL QCD \textit{fitting} method used in \cite{Ikeda:2011bs, Kawanai:2011xb, Allton:2012ki}. In the HAL QCD fitting method, local-extended correlators are fitted to exponentials at large Euclidean time, $\tau$, to extract the Nambu-Bethe-Salpeter (NBS) ground state wavefunction. The NBS wavefunction is then used, in conjunction with the Schr\"{o}dinger equation, to reverse-engineer the potential. The HAL QCD fitting method is understood well from a theoretical point of view since it relies on conventional fitting techniques. However, at non-zero temperature it suffers from familiar limitations --- higher excited states still contribute to the correlator at the largest available $\tau$, making fits unreliable. 

The time-dependent method provides a means to extract the potential from local-extended correlators at moderate $\tau$ and higher temperatures, as described in the following section.
\section{HAL QCD Time-Dependent Approach}
\label{app}
The HAL QCD time-dependent approach takes local-extended correlators as input. Formally, these are constructed from charmonium interpolators,
\begin{align}
J_\Gamma(x;\mathbf{r}) & = \bar{c}(x)\Gamma U(x,x+\mathbf{r})c(x+\mathbf{r}),
\end{align}
where $c(x)$ and $\bar{c}(x)$ are fermion fields, $\Gamma$ is a monomial of gamma matrices and $U(x,x')$ is the product of gauge links which ensures the interpolator's gauge invariance. The local-extended correlator can then be expressed as,
\begin{align}
C_\Gamma(\mathbf{r},\tau) 
& =  \sum_{\mathbf{x}}\langle J_\Gamma(\mathbf{x},\tau;\mathbf{r}) J_\Gamma^\dagger(0;\mathbf{0}) \rangle.
\end{align}
The local-extended correlator can also be expressed as a sum over the eigenstates of the Hamiltonian, $E_j$,
\begin{align}
\label{eq:fullCt}
C_\Gamma(\mathbf{r},\tau) 
&= \sum_j \frac{\psi_j^*(\mathbf{0})\psi_j(\mathbf{r})}{2E_j}\left(e^{-E_j\tau}+e^{-E_j(N_{\tau}-\tau)} \right),
\end{align}
where the $\psi$'s are the NBS wavefunctions at the source and sink. The first step is to consider only the forward-moving contribution to the correlator (the effect of leaving out the backward mover is discussed later),
\begin{align}
\label{eq:forwardCt}
C_\Gamma(\mathbf{r},\tau)=\sum_j \frac{\psi_j^*(\mathbf{0})\psi_j(\mathbf{r})}{2E_j}e^{-E_j\tau}=\sum_j \Psi_j(\mathbf{r})e^{-E_j\tau},
\end{align}
where the $\psi_j^*(\mathbf{0})$ and $2E_j$ have been absorbed into $\Psi_j(\mathbf{r})$ since they are constant for each excited state. The next step is to differentiate both sides w.r.t. $\tau$,
\begin{equation}
\label{eq:wrttau}
\frac{\partial}{\partial \tau}C_\Gamma(\mathbf{r},\tau) = 
- \sum_j E_j \Psi_j(\mathbf{r})e^{-E_j\tau}.
\end{equation}
Then, assuming charm quarks are heavy enough to be treated nonrelativistically the Schr\"odinger equation is applied to $\Psi_j(\mathbf{r})$,
\begin{equation}
\label{eq:schro}
\left(-\frac{\nabla^2}{2\mu} + V_\Gamma(\mathbf{r})\right){\Psi}_j(\mathbf{r})
=
E_j\Psi_j(\mathbf{r}).
\end{equation}
The reduced mass of the charmonium system is defined to be, $\mu=\frac{1}{2}m_c=\frac{1}{4}M_{J/\psi}$, where $m_c$ is the charm mass, and $M_{J/\psi}$ is the vector channel mass.
Using \eqref{eq:schro} in \eqref{eq:wrttau} we obtain,
\begin{align}
\frac{\partial}{\partial \tau}C_\Gamma(\mathbf{r},\tau) = 
\sum_j\left(\frac{\nabla^2}{2\mu}- V_\Gamma(\mathbf{r})\right)\Psi_j(\mathbf{r})e^{-E_j\tau} =
\left(\frac{\nabla^2}{2\mu}- V_\Gamma(\mathbf{r})\right)C_\Gamma(\mathbf{r},\tau).
\end{align}
Finally, this can be rearranged to yield the potential,
\begin{align}
\label{eq:Vt}
V_\Gamma(\mathbf{r})
=
\left(
\frac{\nabla^2C_\Gamma(\mathbf{r},\tau)}{2\mu} 
-
\frac{\partial C_\Gamma(\mathbf{r},\tau)}{\partial \tau}
\right)
\frac{1}{C_\Gamma(\mathbf{r},\tau)}.
\end{align}
The application of \eqref{eq:Vt} has the advantage that the correlators can be used directly to calculate the potential, as opposed to having to fit the correlators at large $\tau$ to extract the ground state NBS wavefunction and then use the Schr\"{o}dinger equation to reverse-engineer the potential. However, note that \eqref{eq:Vt} has an implicit $\tau$ dependence which must be averaged over, see Section \ref{res}. 

In this study only the behaviour of the S-wave potential has been considered. This can be expressed as:
\begin{equation}
V_\Gamma(\mathbf{r}) = V_C(\mathbf{r}) + s_1\cdot s_2V_S(\mathbf{r}),
\end{equation}
where $V_C$ and $V_S$ are the spin-independent and spin-dependent potentials, respectively, and $s_{1,2}$ are the charm quark spins. Knowing the spin product, $s_1\cdot s_2 = -3/4,1/4$, for the pseudoscalar and vector channels, respectively, allows the spin-independent and spin-dependent potentials to be written as linear combinations of the pseudoscalar and vector potentials, $V_{PS}$ and $V_{V}$,
\begin{equation}
\label{eq:Vc}
V_C(\mathbf{r}) = \frac{1}{4}V_{PS}(\mathbf{r}) + \frac{3}{4}V_V(\mathbf{r}),
\end{equation}
\begin{equation}
\label{eq:Vs}
V_S(\mathbf{r}) = V_{V}(\mathbf{r}) - V_{PS}(\mathbf{r}).
\end{equation}
\section{Simulation Details}
\label{simdet}
\begin{table}
\begin{center}
\label{tab:params}
\begin{tabular}{ccccr}
\hline
 \multicolumn{1}{c}{$N_s$} & \multicolumn{1}{c}{$N_\tau$} & \multicolumn{1}{c}{$T$(MeV)} &
\multicolumn{1}{c}{$T/T_c$} & \multicolumn{1}{c}{$N_{\rm cfg}$} \\
\hline 
24 & 40 & 141 & 0.76 & 500 \\      
24 & 36 & 156 & 0.84 & 500 \\ 
24 & 32 & 176 & 0.95 & 1000 \\ 
24 & 28 & 201 & 1.09 & 1000 \\ 
24 & 24 & 235  & 1.27 & 1000 \\ 
\hline
\end{tabular}
\caption{Lattice parameters used, including spatial and temporal
  dimension, $N_s$ and $N_\tau$.\vspace{-1cm}}
\end{center}
\end{table}
The correlator analysis outlined in Section \ref{app} was performed on five different ensembles, equivalent to studying a temperature range of 0.76{\ }-{\ }1.27{\ }T$_\textrm{c}$, where T$_\textrm{c}\approx 185{\ }$MeV. Table\! 1 lists the lattice parameters used. Configurations with 2+1 dynamical flavors of light quarks were generated using a Symanzik gauge action and an anisotropic clover fermion action with stout-smearing \cite{Edwards:2008ja}. The anisotropy of the lattices is $\xi = a_s/a_\tau = 3.5$ with $a_s \simeq 0.12${\ }fm and $a_\tau^{-1} \simeq 5.63${\ }GeV. The charm quark is also simulated with the anisotropic fermion action and its mass is set by tuning the pseudoscalar effective mass to the experimental $\eta_c$ value at zero temperature \cite{Liu:2012ze}. Gaussian smeared sources were employed throughout this study.
\section{Results}
\label{res}

\begin{figure}[b]
\centering
\includegraphics[scale=0.45,trim = 0 0 0 0, clip=true]{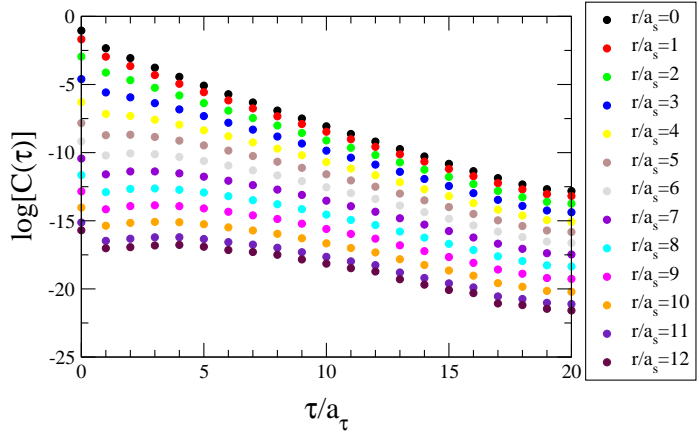}
\caption{Local-extended charmonium correlators for all possible on-axis separations; $N_\tau=40$ ensemble.} 
\label{fig:axisCt}
\end{figure}

In Figure \ref{fig:axisCt} the local-extended charmonium correlators for all possible on-axis separations are plotted for the $N_\tau=40$ ensemble.  As the separation of the charm quarks at the sink increases the signal, being related to the NBS wavefunction, see \eqref{eq:forwardCt}, decreases in magnitude as expected. The correlator for the $r=0$ case is a straight line on a log-plot, even for small $\tau$. For the correlators corresponding to $r\neq 0$, curvature at small $\tau$ is noticeable. This is because the product of the NBS wavefunctions is not positive-definite when the source and sink operators are asymmetric, which means the correlator is not a sum of only positive terms. As higher excited states vanish for larger $\tau$, and the lowest excited state begins to dominate, all the lines become straight.

Figures \ref{fig:40pion}-\ref{fig:28rho} show the result of applying \eqref{eq:Vt} to pseudoscalar and vector channel correlators. Finite differences are taken in the $r$ and $\tau$ directions of data sets like that shown in Figure \ref{fig:axisCt}, to obtain the spatial and temporal derivative terms of \eqref{eq:Vt}. 
The plots have common features: i) The potential values rapidly decrease for the largest $\tau$. To investigate this behaviour, the ground state backward-mover term was added to \eqref{eq:forwardCt}, and the analysis repeated. When the ground state backward-mover term is present, the potential values decrease significantly less rapidly for the largest $\tau$. Therefore, we are confident this feature is due to the absence of the backward-mover term in \eqref{eq:forwardCt}. ii) The $\tau=1$ term is spurious because the finite difference taken in the $\tau$ direction at 
\newpage
\begin{textblock}{20}(-8,0)
\begin{centering}
\begin{figure}
\includegraphics[scale=0.45,trim = 0 0 0 0, clip=true]{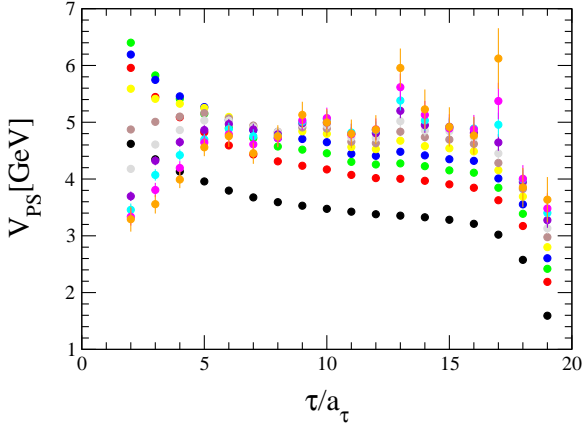}
\caption{Pseudoscalar potential, $V_{PS}$. $N_\tau=40$.}
\label{fig:40pion}
\end{figure}
\end{centering}
\end{textblock}
\begin{textblock}{20}(-1,0)
\begin{centering}
\begin{figure}
\includegraphics[scale=0.45,trim = 0 0 0 0, clip=true]{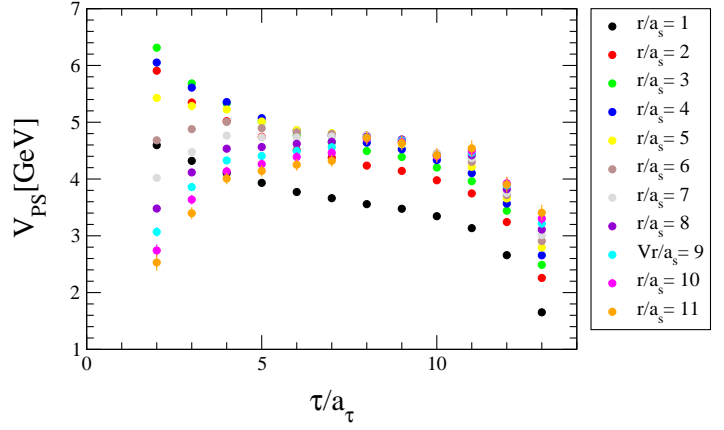}
\caption{Pseudoscalar potential, $V_{PS}$. $N_\tau=28$.}
\label{fig:28pion}
\end{figure}
\end{centering}
\end{textblock}
\begin{textblock}{20}(-8,4)
\begin{centering}
\begin{figure}
\includegraphics[scale=0.45,trim = 0 0 0 0, clip=true]{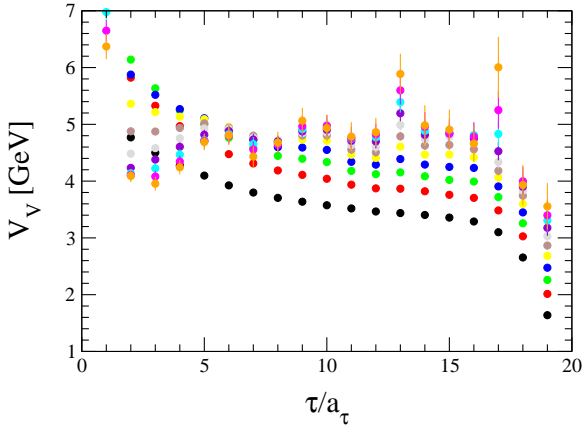}
\caption{Vector potential, $V_V$. $N_\tau=40$.}
\label{fig:40rho}
\end{figure}
\end{centering}
\end{textblock}
\begin{textblock}{20}(-1,4)
\begin{centering}
\begin{figure}
\includegraphics[scale=0.45,trim = 0 0 0 0, clip=true]{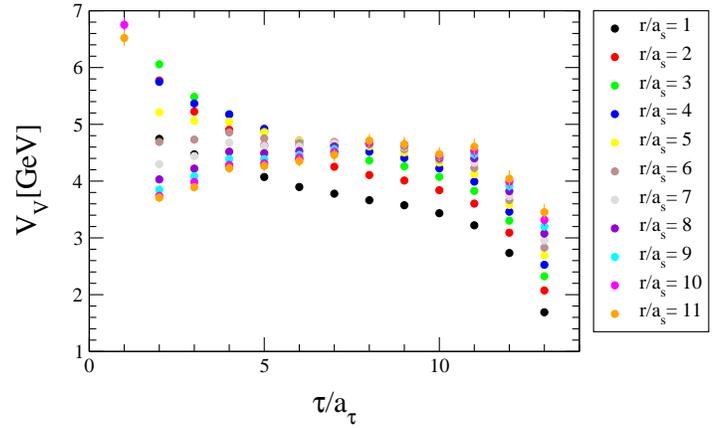}
\caption{Vector potential, $V_V$. $N_\tau=28$.}
\label{fig:28rho}
\end{figure}
\end{centering}
\end{textblock}
\begin{textblock}{20}(-8,8)
\begin{centering}
\begin{figure}
\includegraphics[scale=0.45,trim = 0 0 0 0, clip=true]{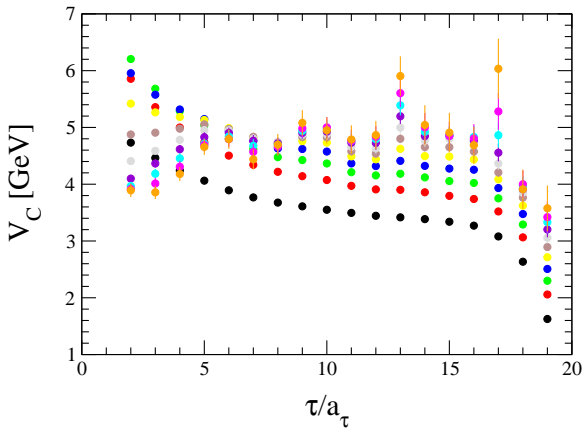}\
\caption{Spin-independent potential, $V_C$. $N_\tau=40$.}
\label{fig:40}
\end{figure}
\end{centering}
\end{textblock}
\begin{textblock}{20}(-1,8)
\begin{centering}
\begin{figure}
\includegraphics[scale=0.45,trim = 0 0 0 0, clip=true]{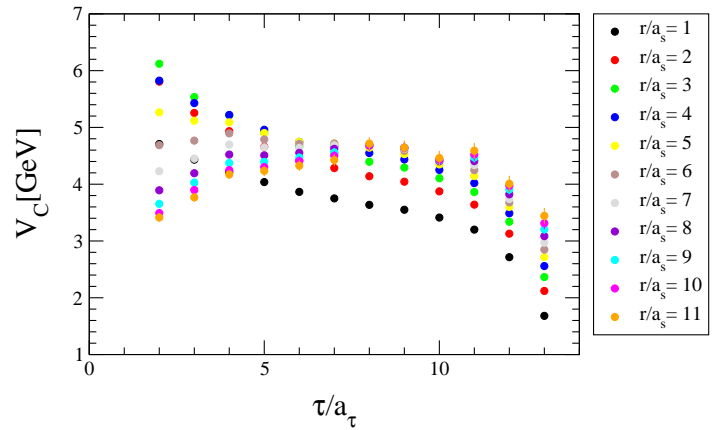}
\caption{Spin-independent potential, $V_C$. $N_\tau=28$.}
\label{fig:28}
\end{figure}
\end{centering}
\end{textblock}
\newpage
\begin{figure}[t]
\begin{minipage}{0.45\linewidth}
\includegraphics[scale=0.45,trim = 0 0 0 0, clip=true]{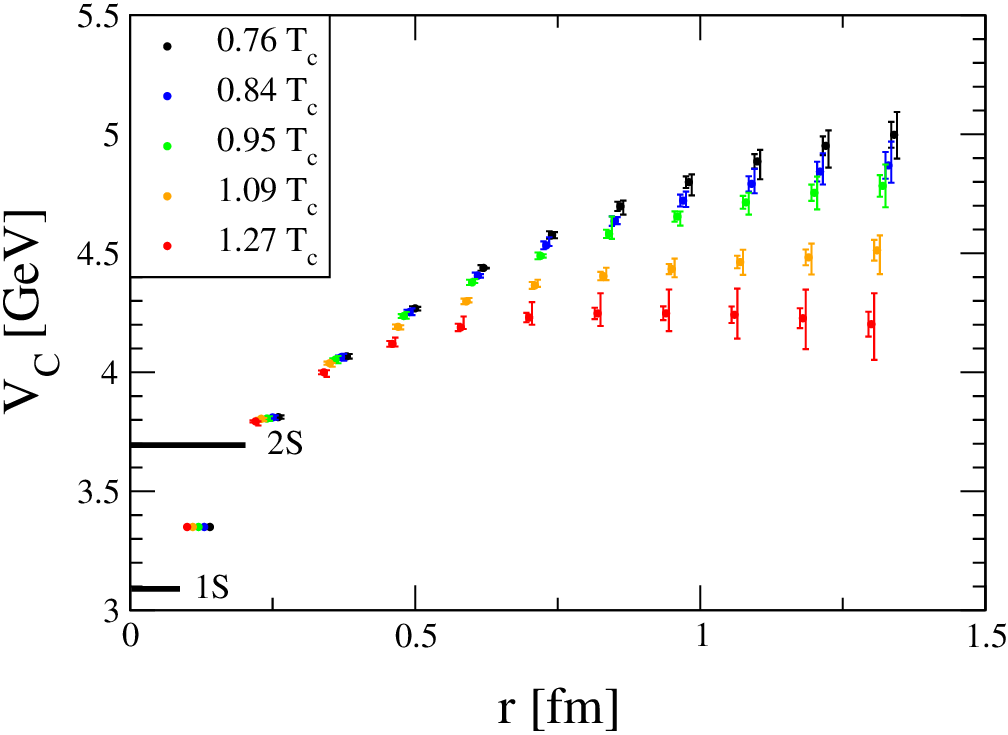} 
\caption{Spin-independent potential, $V_C$, for different $N_\tau$.}
\label{fig:Vc}
\end{minipage}
\hspace{0.5cm}
\quad
\begin{minipage}{0.45\linewidth}
\includegraphics[scale=0.45,trim = 0 0 0 0, clip=true]{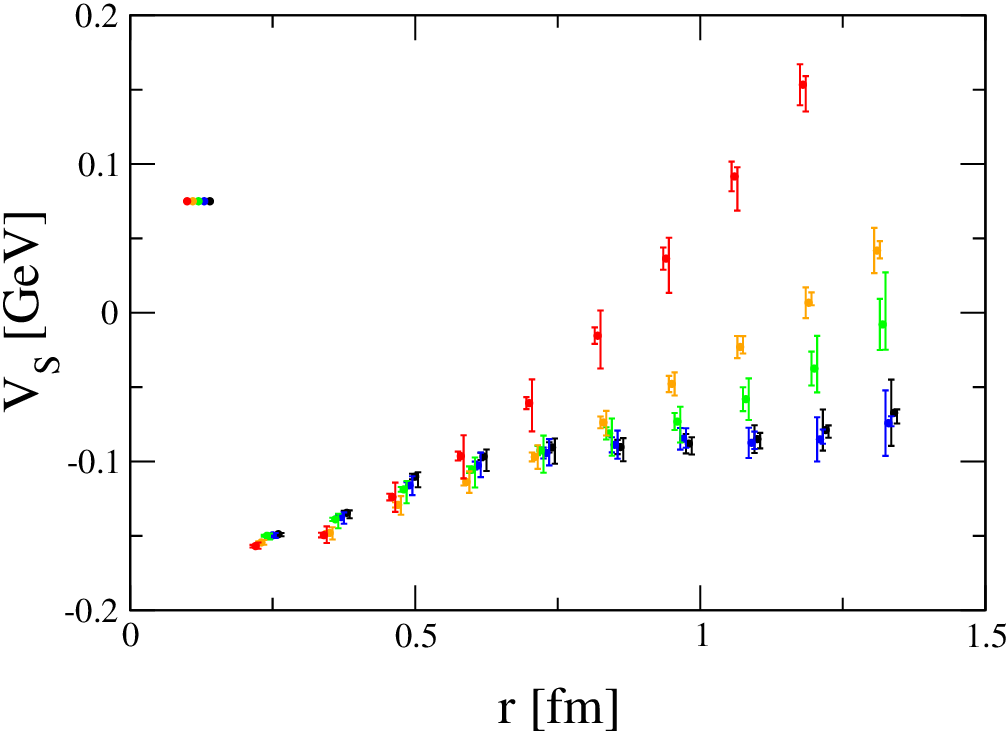} 
\caption{Spin-dependent potential, $V_S$, for different $N_\tau$.}
\label{fig:Vs}
\end{minipage}
\end{figure}

\noindent
this point includes the contact term of the correlator at $\tau=0$. iii) Between $\tau=2$ and $\tau=7$ the potential is not stable. This is thought to be a combination of relativistic effects, lattice artifacts, and the interplay of excited states contributing to the correlator.

Combining the pseudoscalar and vector potentials of Figures \ref{fig:40pion}-\ref{fig:28rho} according to \eqref{eq:Vc} gives the spin-independent plots shown in Figure \ref{fig:40} and Figure \ref{fig:28}. We removed the non-physical $\tau$ dependence in the potential by performing a correlated fit to $V_{C,S}$ to a constant in the $\tau$ interval where it is a plateau: Two other ensembles with $N_\tau=16 \ \& \ 20$ were available, but in these cases the large $\tau$ behaviour associated with the backward-mover and the small $\tau$ behaviour overlap, and there are no reliable plateaus in the potentials. The narrowing of the stable $\tau$ window can be seen by comparing the $N_\tau=40$ and $N_\tau=28$ plots.

Figure \ref{fig:Vc} shows the final result of the analysis for the spin-independent potential. The right-hand error bar represents the systematic uncertainty. This is obtained by varying the start and end of the $\tau$ range within which fits are performed. The left-hand error represents the statistical uncertainty. The masses of the 1S and 2S $J/\psi$ states are included for reference. The potential exhibits a clear temperature dependence, flattening at large $r$ as the temperature increases.

The spin-dependent potential can be calculated by applying \eqref{eq:Vs}, the result is shown in Figure \ref{fig:Vs}. It exhibits a clear repulsive core consistent with the literature \cite{Kawanai:2011jt, Kawanai:2011xb}, and like the spin-independent potential, it also exhibits a temperature dependence.

\section{Conclusions}
\label{con}
There is a significant body of theoretical work studying the interquark potential at non-zero temperature using model, perturbative and lattice (nonperturbative) approaches. However, until now, these lattice studies have all used the static (infinite quark mass) limit. This work improves upon these calculations by considering quarks with finite mass, and thus represents a first-principle calculation of the charmonium potential of QCD at finite temperature. The method we use is based on the HAL QCD time-dependent approach which obtains the potential directly from local-extended correlators. The temperature dependence of the spin-independent charmonium potential is consistent with the expectation that the potential becomes deconfining at high temperature. This work improves upon our earlier work  \cite{Evans:2013yva, Allton:2012ki} in that our lattices are finer and larger volume, and have 2+1 rather than 2 flavors.

\section{Acknowledgements}
\label{ack}
We acknowledge the support and infrastructure provided by the Irish
Centre for High-End Computing, the UK DiRAC Facility jointly funded by STFC,
the Large Facilities Capital Fund of BIS, HPC Wales and Swansea University, and the
PRACE grants 2011040469 and 2012061129. The calculations were carried out
using the Chroma software suite \cite{Edwards:2004sx}. PWME and CA are supported by the STFC. PWME used a Welsh Livery Guild Travel Scholarship to complete this work.   

\bibliographystyle{utphys}
\bibliography{bibliography} 

\end{document}